\documentclass[preprint]{jpsj2}
\usepackage{txfonts}
\title{Phenomenological Theory of a Scalar Electronic Order: \\
Application to Skutterudite PrFe$_4$P$_{12}$}

\author{Annam\'aria \textsc{Kiss}\thanks{E-mail address: amk@cmpt.phys.tohoku.ac.jp} and Yoshio
\textsc{Kuramoto}\thanks{E-mail address: kuramoto@cmpt.phys.tohoku.ac.jp}}

\inst{Department of Physics, Tohoku University, Sendai, 980-8578}

\abst{ 
By phenomenological Landau analysis,
it is shown that a scalar order parameter with the point-group symmetry $\Gamma_{1g}$ explains most properties associated with the phase transition in PrFe$_4$P$_{12}$ at 6.5 K.
The scalar-order model reproduces
magnetic and elastic properties 
in PrFe$_4$P$_{12}$ consistently such as (i) the anomaly of the magnetic susceptibility and elastic constant at the transition temperature, (ii) anisotropy of the magnetic susceptibility in the presence of uniaxial pressure, and (iii) the anomaly in the elastic constant in magnetic field.
An Ehrenfest relation is derived which relates the anomaly of the magnetic susceptibility to that of the elastic constant at the transition.
}

\kword{skutterudite, PrFe$_4$P$_{12}$, scalar order, magnetic susceptibility, elastic constant, uniaxial pressure}

\begin{document}
\maketitle

\section{Introduction}

Rare earth filled skutterudites RT$_{4}$X$_{12}$ attract continuous interest because of their intriguing and rich variety of phenomena depending on the rare earth $R$.  
Owing to intensive experimental and theoretical efforts,
some of these phenomena are now understood and explained, but there still remain unsolved problems of great significance.  
Among them are mysterious ordered phases in PrFe$_4$P$_{12}$ \cite{aoki,tayama}, metal-insulator transition and magnetic transitions under high pressure \cite{hidaka}, and ferromagnetic ordered phase when a few percent of the Pr concentration is substituted by La\cite{namiki}.
The purpose of this paper is to provide a coherent phenomenological picture for  the fundamental order in PrFe$_4$P$_{12}$, in terms of a scalar electronic order.
Other ordered phases in PrFe$_4$P$_{12}$ have different order parameters, and
require separate analysis.

PrFe$_{4}$P$_{12}$ undergoes a phase transition at $T_{0}=6.5$K, which accompanies a sharp peak in the magnetic susceptibility \cite{aoki}.
Staggered dipole moments are found in the ordered phase in the presence of a magnetic field.  This observation indicates that the order parameter does not break the time-reversal symmetry \cite{hao2}.
A modulation with the wave vector ${\bf Q}=(1,0,0)$ was found below $T_{0}$ by X-ray diffraction experiments \cite{iwasa3}, which is attributed to the existence of staggered local electronic states of the Pr ions.
Early NMR \cite{kikuchi2} and elastic measurements \cite{nakanishi1} were interpreted in terms of antiferro ordering of $\Gamma_{3}$-type quadrupole moments, which picture was widely accepted for a long time.
However, 
the isotropy of the magnetic susceptibility for different field directions even in the ordered phase cannot be explained by $\Gamma_{3}$ quadrupolar order. Furthermore, staggered dipoles perpendicular to the field direction are found neither in neutron diffraction \cite{hao2, hao1} nor NMR \cite{kikuchi1}.
Recently, careful analysis of the NMR results pointed out that the local symmetry at the Pr sites is preserved in the ordered phase \cite{kikuchi1, sakai}. 
Furthermore, the continuous field-angle dependence of the transition temperature gives also evidence for the exclusion of $\Gamma_{3}$ quadrupolar order \cite{sakakibara}.

In a previous paper  \cite{utolso}, we proposed that the order parameter in 
PrFe$_4$P$_{12}$ is a staggered electronic order which does not break the local $T_h$ symmetry around each Pr site.  We call this order with the $\Gamma_{1g}$ symmetry a scalar order.   
It is found in ref.~\citen{utolso} that the scalar order scenario can explain naturally the isotropic magnetic susceptibility in the ordered phase, the field angle dependence of the transition temperature and magnetization, and also the splitting pattern of the $^{31}$P NMR spectra.
The main purpose of this paper is to provide a coherent description of available 
magneto-elastic properties of PrFe$_4$P$_{12}$ within the scalar order scenario.
We can explain not only elastic anomalies in PrFe$_4$P$_{12}$ near the phase transition, but also the huge anisotropy of the magnetic susceptibility under uniaxial pressure.   Our model assumes phenomenological
couplings of the scalar, dipolar, and quadrupolar degrees of freedom.

This paper is organized as follows. 
In section 2 we introduce the free energy expansion around the transition temperature and clarify its constituents. 
Section 3 discusses the properties of magnetic susceptibility, and its anisotropy in the presence of uniaxial pressure is obtained. In addition,
particular attention is paid to the emergent anisotropy in magnetic field.
In section 4 we discuss elastic properties of the scalar order within the phenomenological framework.
Section 5 studies Ehrenfest relations relating to the anomalies of magnetic and elastic quantities.
The last section is devoted to the summary of the paper.

\section{General framework of phenomenology}
\subsection{Landau expansion of the free energy}\label{sec:landau}

A scalar order can be two different kinds: one which breaks the time reversal symmetry ($\Gamma_{1u}$), and the other which does not ($\Gamma_{1g}$).
In this paper we assume a two-sublattice scalar order with the ordering vector ${\bf Q}=(1,0,0)$, where the order parameter has the $\Gamma_{1g}$ symmetry.

In the Landau theory, one expands the free energy in terms of the set of electronic order parameters $\Psi_i$, which are taken to be real.  
Up to fourth-order, we write
\begin{align}
{\cal F}(\Psi) &=
\sum_i\left( 
\frac{1}{2} \alpha_i \Psi_i^2 + 
\frac{1}{4}b_{i}\Psi_i^4  \right)+ 
\sum_{i\neq j} \left( 
g_{ij}\Psi_i^2 \Psi_j +
\frac 12 c_{ij}\Psi_i^2 \Psi_j^2
 \right),
 \end{align}
where 
we have introduced the quantity $\alpha_i = a_i (T-T_i)$.

The constants
$a_i, b_i$ 
are positive, while $g_{ij}$ and $c_{ij}$ can have either sign.
$T_i$ is a hypothetical transition temperature without coupling to other order parameters.   The actual transition occurs at 
$T_0$ corresponding to the scalar order, which gives the largest of all $T_i$.
For other component $\Psi_i$, we neglect the corresponding $b_i$ in most cases.
These parameterizations have a merit that each coefficient is regarded as a constant as long as the temperature is close to $T_0$, and
the external perturbations are small.

As explicit constituents of $\Psi_i$, 
we include the scalar order parameter $\psi_{\bf Q}$, the homogeneous magnetization $\vec{M}$, and the $\Gamma_3$-type homogeneous quadrupoles.
Furthermore, we also include the lattice strain components $\varepsilon_{xx}$, $\varepsilon_{yy}$, $\varepsilon_{zz}$,
which have a bilinear coupling with quadrupole moments $Q_{ij}$.
The second-order couplings $c_{i\varepsilon}$ with $i=Q,\psi_Q$
can be neglected because the background elastic constant $C_{ij}^{(0)}$ is large enough.
The external magnetic field is assumed with general direction, and the magnetization components are included up to fourth order. 
We write the free energy 
consistent with the cubic symmetry as
\begin{eqnarray}
{\cal F}(\psi_{\bf Q}, Q, \vec{M}, \varepsilon) &=&  {\cal F}_l+ \frac{1}{2} \alpha_{\psi} \psi_{\bf Q}^2 + \frac{1}{4}b_{\psi}\psi_{\bf Q}^4  
+ \frac{1}{2}\alpha_{M} \vec{M}^2+
\frac{1}{2}c_{\psi M}\psi_{\bf Q}^2  \vec{M}^2
 \nonumber\\
 &+& \frac{1}{2}\alpha_{Q}Q^2 +  \frac{1}{2}
c_{ M Q} \vec{M}^2 Q^2  + \frac{1}{2}c_{\psi Q}\psi_{\bf Q}^2 Q^2 
+ g_{\psi \varepsilon}\psi_{\bf Q}^2\varepsilon_s
\nonumber\\
&+&   
g_{Q \varepsilon} Q^2\varepsilon_s 
+ g_{M \varepsilon}\vec{M}^2  \varepsilon_{s} + {\cal O}(M^4) + {\cal O}(Q^4)
  \nonumber\\
 &+&  
 g_{M Q}\left[ \frac{1}{\sqrt{6}} \left(2M_{z}^2 -M_{x}^2-M_{y}^2\right)Q_{u}+\frac{1}{\sqrt{2}} \left(M_{x}^2-M_{y}^2 \right)Q_{v}\right],
 \label{eq:gibbs3}\\
 {\cal F}_l
&=& B \left(\varepsilon_{u} Q_{u}+\varepsilon_{v} Q_{v}\right)
+   
\frac{1}{2}\left(C_{11}^{(0)}-C_{12}^{(0)}\right) \left(\varepsilon_{u}^2+\varepsilon_{v}^2\right)
 + \frac{1}{2}\left(C_{11}^{(0)}+2C_{12}^{(0)}\right)\varepsilon_s^2,
\label{eq:gibbs3a}
\end{eqnarray} 
where $\alpha_{\psi}=a_\psi (T-T_0)$, $\alpha_{Q}=a_Q (T-T_Q)$, $\alpha_{M}=a_M (T-T_F)$ and $\vec{M}^2=M_{x}^2+M_{y}^2+M_{z}^2$.
For the strain components we introduce the notations
$\varepsilon_{u} =(1/\sqrt{6})(2\varepsilon_{zz}-\varepsilon_{xx}-\varepsilon_{yy})$, $\varepsilon_{v} =(1/\sqrt{2})(\varepsilon_{xx}-\varepsilon_{yy})$ and $\varepsilon_s=(1/\sqrt{3})(\varepsilon_{xx}+\varepsilon_{yy}+\varepsilon_{zz})$, and $C_{11}^{(0)}$ and $C_{12}^{(0)}$ are the elastic constants in the background. $Q_{u}$ and $Q_{v}$ are the $\Gamma_{3}$ quadrupole moments defined as $Q_{u}= O_{2}^{0}=(1/\sqrt{6})(2J_{z}^2-J_{x}^2-J_y^2)$, $Q_{v}= O_{2}^{2} = (1/\sqrt{2})(J_{x}^2-J_y^2)$, and we introduced the notation $Q^2=Q_{u}^2+Q_{v}^2$.

As external fields, we consider only two cases: magnetic field and uniaxial stress.
We will consider the uniaxial stress direction $\sigma\|(001)$, so that  
we have $\sigma_{zz}\ne 0$ and $\sigma_{xx}=\sigma_{yy}=0$. This stress can be decomposed into the sum of an isotropic and a traceless part as
\begin{eqnarray} 
 \left(
\begin{array}{c}
          \sigma_{xx}\\
           \sigma_{yy}\\
           \sigma_{zz}\end{array}
       \right)=
 \left(
\begin{array}{c}
          0\\
           0\\
           \sigma\end{array}
       \right)=
 \frac{1}{3} \sigma \left[ \left(
\begin{array}{c}
          1\\
           1\\
          1\end{array}
       \right) +   \left(
\begin{array}{c}
          -1\\
           -1\\
           2\end{array}
       \right)\right].
\label{eq:smatrixu2}
\end{eqnarray}
Here, the first part $(1,1,1)$ has $\Gamma_{1}$ symmetry and it is the isotropic hydrostatic pressure component. The second part $(-1,-1,2)$ has $\Gamma_{3}$ symmetry, and it causes the lowering of the original cubic symmetry to tetragonal one.
For the uniaxial stress components we introduce the notations 
$\sigma_{u} =(1/\sqrt{6})(2\sigma_{zz}-\sigma_{xx}-\sigma_{yy})$, $\sigma_{v} =(1/\sqrt{2})(\sigma_{xx}-\sigma_{yy})$ and $\sigma_s=(1/\sqrt{3})(\sigma_{xx}+\sigma_{yy}+\sigma_{zz})$, where we have the constraint $\sigma_u=\sqrt{2}\sigma_s$ and $\sigma_{v}=0$.

The equilibrium condition leads to the coupled equations:
\begin{align}
H_{k} &=\frac{\partial\cal F}{\partial M}_{k} = \tilde{\alpha}_{M,k} M_{k},
&\sigma_s &=\frac{\partial\cal F}{\partial \varepsilon_s} =  C_0 \varepsilon_s +g_{\psi\varepsilon}\psi_{\bf Q}^2
+g_{M \varepsilon}\vec{M}^2 +g_{Q \varepsilon}Q^2,\nonumber
\\
\sigma_u &=\frac{\partial\cal F}{\partial \varepsilon_u} = C_3 \varepsilon_u+BQ_u,
&0&=\frac{\partial\cal F}{\partial Q_u} = \tilde{\alpha}_Q Q_u+B\varepsilon_u +
 \frac{1}{\sqrt{6}} g_{M Q} \left(2M_{z}^2-M_{x}^2-M_{y}^2 \right),\nonumber\\
0 &=\frac{\partial\cal F}{\partial \varepsilon_v} = C_3 \varepsilon_v+BQ_v,
&0&=\frac{\partial\cal F}{\partial Q_v} = \tilde{\alpha}_Q Q_v+B\varepsilon_v +\frac{1}{\sqrt{2}} g_{M Q} \left(M_{x}^2-M_{y}^2 \right),\nonumber\\
0&=\frac{\partial\cal F}{\partial \psi_{\bf Q}} = \tilde{\alpha}_\psi \psi_{\bf Q},\label{eqcon}
\end{align}
where $k=x,y,z$, and $\tilde{\alpha}_i$ indicates a renormalized inverse susceptibility.
They are given together with $C_3$ and $C_0$ by
\begin{align}
\tilde{\alpha}_{M,x} &= \alpha_M+ 2g_{M\varepsilon}\varepsilon_s+c_{\psi M}\psi_{\bf Q}^2 + c_{M Q}Q^2+g_{MQ}\frac{2}{\sqrt{6}}\left(\sqrt{3}Q_{v}-Q_u\right), \\
\tilde{\alpha}_{M,y} &= \alpha_M + 2g_{M\varepsilon}\varepsilon_s+c_{\psi M}\psi_{\bf Q}^2 + c_{M Q}Q^2+g_{MQ} \frac{2}{\sqrt{6}} \left(-\sqrt{3}Q_{v}-Q_u\right), \\
\tilde{\alpha}_{M,z} &= \alpha_M + 2g_{M\varepsilon}\varepsilon_s+c_{\psi M}\psi_{\bf Q}^2 + c_{M Q}Q^2+g_{MQ} \frac{4}{\sqrt{6}} Q_u, \\
\tilde{\alpha}_Q &= \alpha_Q+
2g_{Q\varepsilon}\varepsilon_s+c_{\psi Q}\psi_{\bf Q}^2 +c_{ M Q}\vec{M}^2, \\
\tilde{\alpha}_\psi &= \alpha_\psi +
2g_{\psi \varepsilon}\varepsilon_s+ b_\psi \psi_{\bf Q}^2+c_{\psi Q} Q^2+c_{\psi M}\vec{M}^2, 
\label{tilde_alpha} \\
C_3 &= C_{11}^{(0)}-C_{12}^{(0)},\\
C_0 &= C_{11}^{(0)}+2C_{12}^{(0)}. 
\end{align}

\subsection{Phase boundary}

The transition temperature $T_c=T_0+\Delta T_0$ is obtained by the condition $\tilde{\alpha}_\psi=0$ in eq.(\ref{tilde_alpha}).
Let us first consider the 
case under magnetic field, but with zero stress. We assume that the magnetic field is parallel to the direction $(001)$, which gives $M_{x}=M_{y}=0$, and we write $M_{z}\equiv M$. We note that the final result does not depend on the field direction up to second order in the field.
The equilibrium condition is given by
\begin{align}
H  &=   \alpha_M M, 
\label{F/M}\\
0  &=   C_{0}\varepsilon_{s}+g_{M \varepsilon}M^2, 
\label{F/es}\\
0 & =   \tilde{\alpha}_\psi \psi_{\bf Q}=(\alpha_\psi +c_{\psi M}  M^2+2g_{\psi \varepsilon}\varepsilon_{s}) \psi_{\bf Q}. 
\label{F/psi}
\end{align}
The condition $\tilde{\alpha}_\psi =0$ gives
the temperature-magnetic field phase boundary.
By eliminating $M$ and $\varepsilon_{s}$ in eq.(\ref{F/psi}) with the aid of eqs.(\ref{F/M}) and (\ref{F/es}), we obtain 
\begin{eqnarray}
T_c
= T_{0} -\frac{1}{a_{\psi} a_M^2 (T_{0}-T_F)^2}\left( c_{\psi M}-\frac{2g_{\psi \varepsilon}g_{M \varepsilon}}{C_{0}} \right)H^2 \equiv T_{0} -t_{H2}H^2.\label{pbh2}
\end{eqnarray}
We will neglect the terms where the background elastic constants $C_{0}$ or $C_{3}$ appear in the denominator since these terms are small. The exception arises when these terms are multiplied by the uniaxial pressure, because the product $(1/C_{0})\sigma$ is not small.
Thus, we obtain from expression (\ref{pbh2})
\begin{eqnarray}
T_c= T_{0} -\frac{c_{\psi M}H^2}{a_{\psi} a_M^2 (T_{0}-T_F)^2}.\label{pbh2p}
\end{eqnarray}
From the measured temperature-magnetic field phase boundary we obtain $t_{H2}\approx 0.34$[K$\cdot$T$^{-2}$].\cite{aoki} 
Therefore, we estimate $c_{\psi M}/[a_{\psi}a_M^2 (T_{0}-T_F)^2]\approx 0.34$[K$\cdot$T$^{-2}$].

\begin{figure}
\centering
\includegraphics[height=5cm,angle=0]{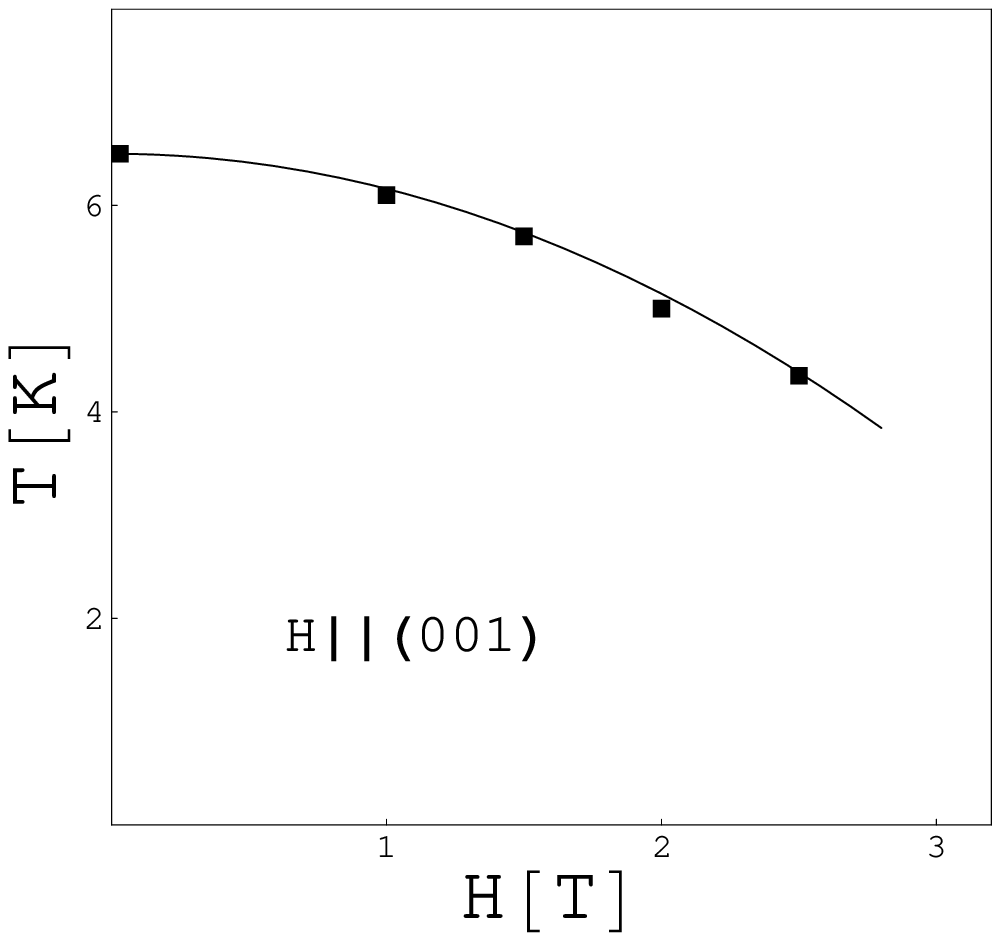}\hspace*{1cm}
\includegraphics[height=5cm,angle=0]{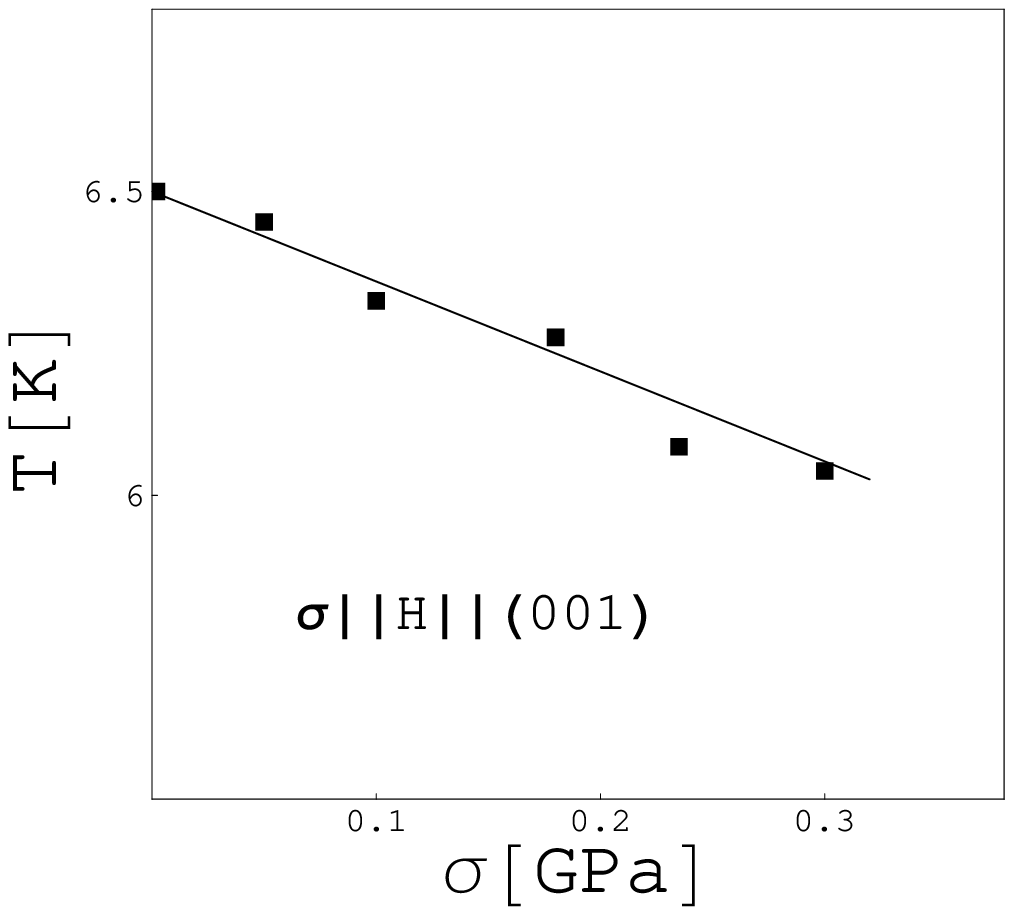}
\hspace*{1cm}
\caption{Temperature-magnetic field ({\sl left}) and temperature-uniaxial pressure ({\sl right}) phase boundaries close to the transition temperature $T_{0}=6.5$K. For the parameter values, see the text.
Black boxes represent the measured result taken from refs.~\citen{aoki} and \citen{saha}.}\label{fig:7}
\end{figure}

The phase boundary under uniaxial stress $\sigma\ne 0$
can be obtained in a similar manner. 
The experimental temperature-uniaxial pressure phase boundary\cite{saha} is linear for low values of $\sigma$.
Hence we concentrate on the linear regime, where only the scalar combination  $\sigma_s = 1/\sqrt{3} (
\sigma_{xx}+\sigma_{yy}+\sigma_{zz})
$ 
enters. 
Without magnetic field, we obtain 
\begin{align}
T_c = T_{0}- \frac{2g_{\psi \varepsilon}}
{a_{\psi} C_{0}} \sigma_{s} \equiv T_{0}
-t_{\sigma 1} \sigma_{s}.\label{pbs1}
\end{align}
The estimate
 $t_{\sigma 1} /\sqrt{3}\approx 1.5$[K$\cdot$GPa$^{-1}$]\cite{saha} gives the restriction for the combination: 
$g_{\psi \varepsilon}/(a_{\psi} C_{0})
\approx 1.3$[K$\cdot$GPa$^{-1}$].
The fitting result for the $T$-$H$ and $T$-$\sigma$ phase boundaries are shown in Fig.~\ref{fig:7}.

For later convenience in discussing Ehrenfest relations, we also go
on to the second order regime, where two different contributions enter: $\sigma_s^2 = 1/3 (
\sigma_{xx}+\sigma_{yy}+\sigma_{zz})^2$ and $\sigma_{u}^2+\sigma_{v}^2= 2/3 (\sigma_{xx}^2+\sigma_{yy}^2+\sigma_{zz}^2-\sigma_{xx}\sigma_{yy}-\sigma_{xx}\sigma_{zz}-\sigma_{yy}\sigma_{zz})$.
Since we neglected the second-order coupling $c_{\psi \varepsilon}$ in the free energy expansion (\ref{eq:gibbs3}), the term with $(
\sigma_{xx}+\sigma_{yy}+\sigma_{zz})^2$ cannot be calculated. For the other second order term we obtain
\begin{eqnarray}
 -\frac{1}{a_{\psi}} \frac{B^2}{[B^2-a_{Q}(T_{0}-T_{Q})]^2}\left( c_{\psi Q}-\frac{2g_{\psi \varepsilon}g_{Q \varepsilon} }{C_{0}}\right)\left( \sigma_{u}^2+\sigma_{v}^2\right)\equiv -t_{\sigma 2}\left( \sigma_{u}^2+\sigma_{v}^2\right).\label{pbs2}
\end{eqnarray}

\section{Magnetic susceptibility}

\subsection{No uniaxial stress}

In the case of zero uniaxial stress, the magnetic susceptibility is obtained from eq.(\ref{eq:gibbs3}) by setting $\sigma_{u}=\sigma_{s}=0$ in the equilibrium condition (\ref{eqcon}). For $T>T_0$, we further set $\psi_{\bf Q}=0$, and 
obtain the Curie-Weiss law:
\begin{align}
\chi_{+}^{-1}= \frac{\partial^2\cal F}{\partial M^2} = a_M (T-T_F),
\label{chi+}
\end{align}
where $T_{F}$ is the Weiss temperature, and we use the superscript + to indicate $T>T_0$.
In the ordered phase close to the transition temperature, we 
obtain 
\begin{eqnarray}
\chi_{-}^{-1} &=& a_{M}(T-T_{F}) +c_{\psi M} \psi_{\bf Q}^2 +2g_{M \varepsilon}\varepsilon_{s}\nonumber\\
&=&
a_{M}(T-T_{F}) -\frac{c_{\psi M} a_\psi (T-T_0)}{b_{\psi}}+\frac{2a_\psi (T-T_0)g_{\psi \varepsilon}}{b_{\psi}C_{0}-2g_{\psi \varepsilon}^2}\left(g_{M \varepsilon}-\frac{c_{\psi M}g_{\psi \varepsilon}}{b_{\psi}} \right),\label{fullsus}
\end{eqnarray}
where we set $\psi_{\bf Q}^2$ and $\varepsilon_{s}$
using the equilibrium conditions 
$\partial {\cal F}/\partial \psi_{\bf Q} =0$ and $\partial {\cal F}/\partial \varepsilon_{s} =0$.
Neglecting the last term in eq.(\ref{fullsus}), we obtain
the inverse magnetic susceptibility as
\begin{eqnarray}
\chi_{-}^{-1} = 
a_{M}(T-T_{F}) -\frac{c_{\psi M} a_\psi (T-T_0)}{b_{\psi}}.\label{chi-}
\end{eqnarray}
There is a peak in the magnetic susceptibility at $T=T_0$ if $\partial \chi_{-}^{-1}/\partial T|_{T_{0}}<0$ is satisfied.
The temperature derivative is calculated as
\begin{eqnarray}
\left. \frac{\partial \chi_{-}^{-1}}{\partial T}\right|_{T_{0}}=a_{M}-
\frac{
a_{\psi} c_{\psi M}}{b_\psi}, 
\end{eqnarray}
which gives the condition $a_{\psi} c_{\psi M}>a_{M}b_{\psi}$
for occurrence of the peak in $\chi (T)$ at $T=T_0$.
Expressions (\ref{chi+}) and (\ref{chi-})
for the magnetic susceptibility are used to fit the measured result. 
We set $T_{F}=3.5$K for the Weiss temperature, which is found in the experiments \cite{aoki}.
The value for the parameter $a_{M}$ is obtained from the fit in the paramagnetic phase
as $a_{M}=9.6\cdot 10^{3}$[Pa$\cdot$$\mu_{B}^{-2}$$\cdot$K$^{-1}$]. Additionally, the combination $a_{\psi}c_{\psi M}/b_{\psi}$ is obtained from the fit of the susceptibility in the ordered phase. We can choose the value of $b_{\psi}$ and then our previous estimation for the parameter $t_{H2}$ from the measured temperature-magnetic field phase boundary gives the values of $a_{\psi}$ and $c_{\psi M}$.
The results are given by $b_{\psi}=10^{4}$[Pa], $a_{\psi}=1.95\cdot 10^{4}$[Pa$\cdot$K$^{-1}$], $c_{\psi M}=1.37\cdot 10^{4}$[Pa $\cdot$$\mu_{B}^{-2}$].
The fitting result obtained is shown in the left part of Fig.~\ref{fig:4}.

\subsection{Finite uniaxial stress}

Intriguing experimental results are obtained for the magnetic susceptibility in the presence of uniaxial stress \cite{saha,matsuda}. Namely, 
the magnetic susceptibility shows large enhancement for the uniaxial pressure applied parallel to the magnetic field direction ($H\|\sigma$), 
while it shows only slight decrease when the pressure is applied perpendicular to the field direction ($H\perp\sigma$).

Now we discuss the properties of the magnetic susceptibility around the transition temperature $T_{0}$.
The direction of the uniaxial stress is taken as
$\sigma\|(001)$, and we consider two different directions of the magnetic field, namely
$H\|(001)$ and $H\|(100)$. These two cases can be obtained from
the free energy given by eq.(\ref{eq:gibbs3})
by taking 
$M_{z}\ne 0$ and $M_{x}\ne 0$, respectively.
For small values of the uniaxial stress, it is enough to consider only the linear term in $\sigma$.
Thus,
we obtain the susceptibilities 
$\chi_\parallel$ for $H\|(001)$ and 
$\chi_\perp$ for $H\|(100)$ as 
\begin{align}
\chi_\parallel^{-1} = 
\alpha_M+\frac{4}{\sqrt{6}}g_{MQ}Q_{u}+ 
2g_{M\varepsilon}\varepsilon_s+c_{\psi M}\psi_{\bf Q}^2,\label{psuspar} \\
\chi_\perp^{-1} = 
\alpha_M- \frac{2}{\sqrt{6}} g_{MQ}Q_{u}+ 
2g_{M\varepsilon}\varepsilon_s+c_{\psi M}\psi_{\bf Q}^2. \label{psusperp}
\end{align}
Among the terms appearing in the expression of the magnetic susceptibility, $g_{M\varepsilon}\varepsilon_s$ and $g_{MQ}Q_{u}$ contain the uniaxial stress.
We can find in eqs.(\ref{psuspar}) and (\ref{psusperp}) that the former term is isotropic, while the latter term is anisotropic and has different sign for the two magnetic field directions.
Therefore, the cancellation of the isotropic and anisotropic terms can occur for the case $H\perp\sigma$, which reproduces the experimental behavior.
\begin{figure}
\centering
\includegraphics[totalheight=5cm,angle=0]{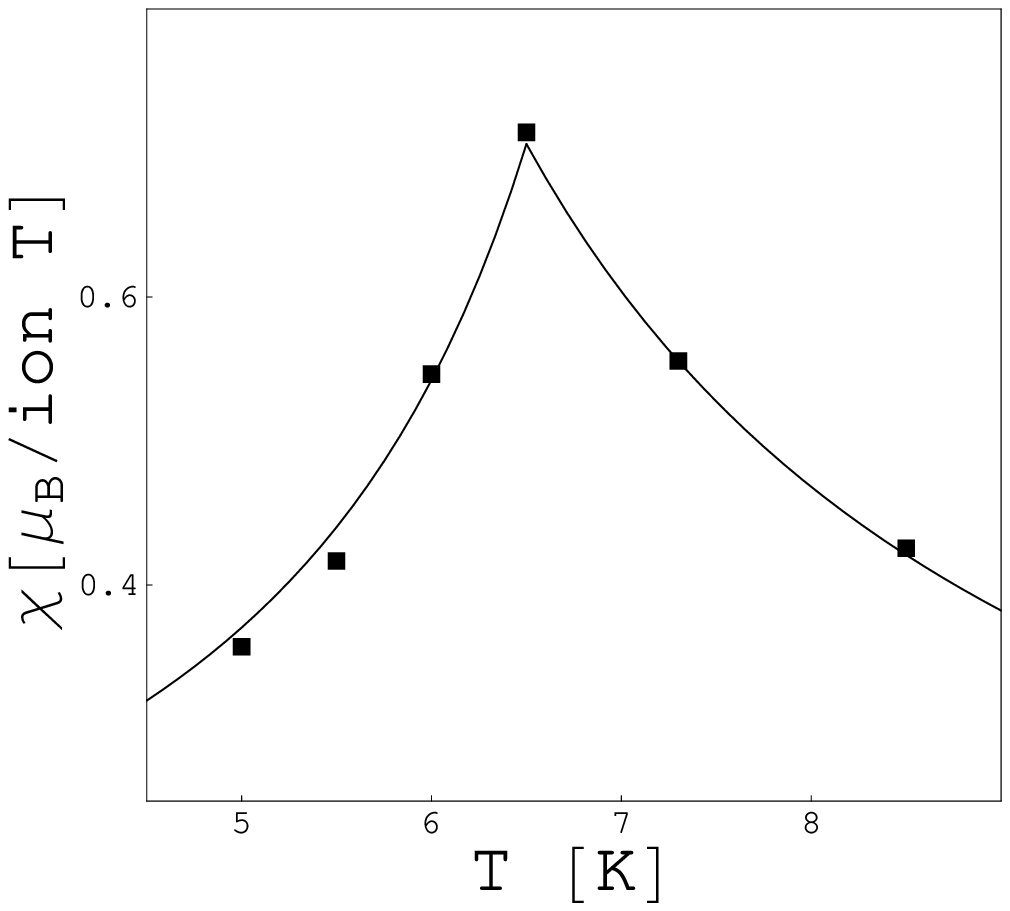}\hspace*{1cm}
\includegraphics[totalheight=5cm,angle=0]{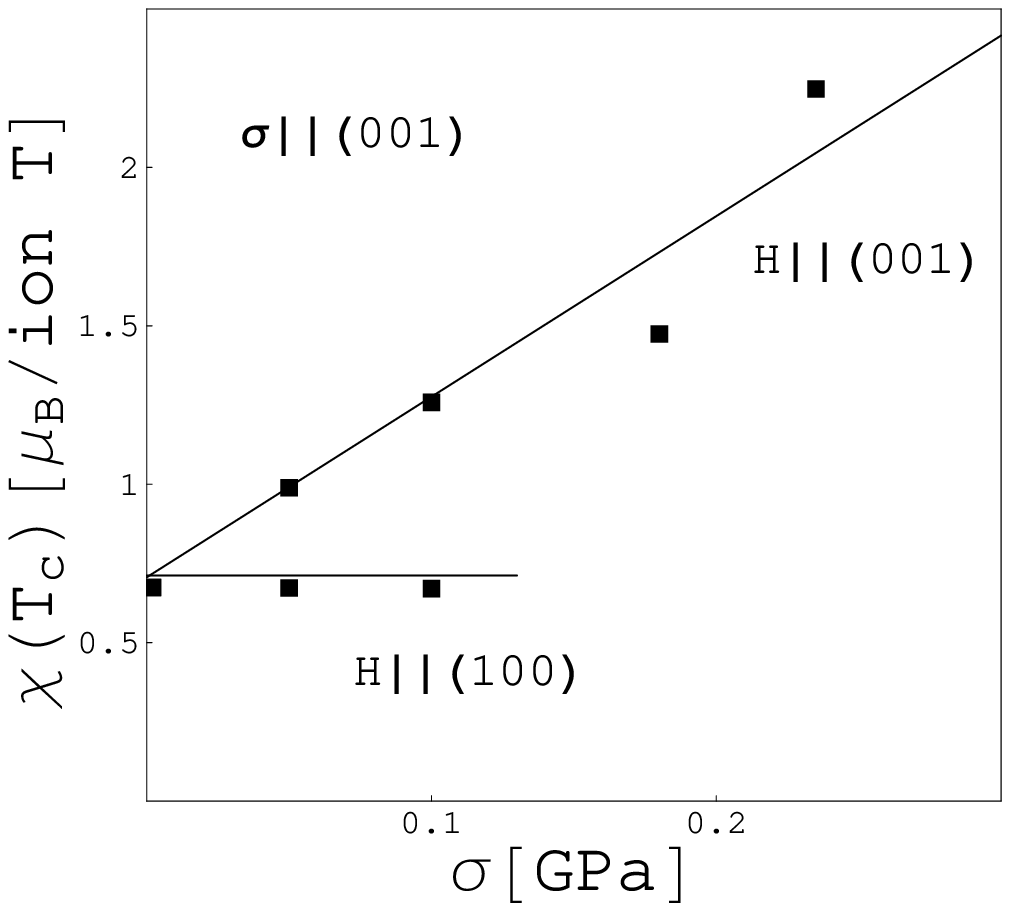}
\caption{{\sl Left:} Magnetic susceptibility 
around the transition temperature $T_0$. 
Black boxes represent the measured result taken from ref.~\citen{aoki}.
{\sl Right:} Uniaxial pressure dependence of  the susceptibility peak at $T=T_{c}$ for the magnetic field directions $(001)$ and $(100)$. Black boxes represent the measured result taken from ref.~\citen{saha}, where we converted the magnetic susceptibility from unit of [emu$\cdot$mol$^{-1}$] to [$\mu_{B}\cdot$ion$^{-1}$ T$^{-1}$]. 
}\label{fig:4}
\end{figure}

Now we set $\psi_{\bf Q}$, $\varepsilon_{s}$ and $Q_{u}$ from the equilibrium conditions 
$\partial {\cal F}/\partial \psi_{\bf Q} =0$, $\partial {\cal F}/\partial \varepsilon_{s} =\sigma_{s}=\sqrt{1/3}\sigma$ and $\partial {\cal F}/\partial \varepsilon_{u} =\sigma_{u}=\sqrt{2/3}\sigma$. 
Keeping only the leading term as we did also previously in the susceptibility calculation with no uniaxial stress, 
we obtain the magnetic susceptibilities $\chi_\parallel$ and $\chi_\perp$ in the ordered phase as
\begin{eqnarray}
\chi_{\parallel,-}^{-1} &=&  \chi_{-}^{-1} - \frac{4}{3} \left( \frac{B g_{M Q}}{C_{3}\tilde{\alpha}_{Q}-B^2}\right) \sigma + \frac{2}{\sqrt{3}} \left(\frac{g_{M \varepsilon}}{C_{0}}\right) \sigma,\label{ssuscz}\\
\chi_{\perp,-}^{-1} &=& \chi_{-}^{-1}+ \frac{2}{3} \left(\frac{B g_{M Q}}{C_{3}\tilde{\alpha}_{Q}-B^2} \right) \sigma + \frac{2}{\sqrt{3}} \left(\frac{g_{M \varepsilon}}{C_{0}}\right) \sigma,\label{ssuscx}
\end{eqnarray}
where $\chi_{-}$ is given by eq.(\ref{chi-}), and
\begin{eqnarray}
\tilde{\alpha}_{Q}=\alpha_{Q}-\frac{c_{\psi Q}a_\psi (T-T_0)}{b_{\psi}} .
\end{eqnarray}
The susceptibilities in the paramagnetic phase can be obtained by taking $\alpha_{Q}=\tilde{\alpha}_{Q}$ in eqs.(\ref{ssuscz}) and (\ref{ssuscx}).
In experiment, the uniaxial pressure dependence of the magnetic susceptibility at the transition temperature\cite{saha} is linear for small values of $\sigma$. We make series expansion in terms of $\sigma$ in the expressions (\ref{ssuscz}) and (\ref{ssuscx}), and obtain the susceptibilities $\chi_\parallel$ and $\chi_\perp$ at the transition temperature $T_{c}$ as
\begin{eqnarray}
\chi_\parallel(T_{c}) &\approx&  \frac{1}{a_M (T_c-T_F)} - \frac{2}{\sqrt{3}}\frac{1}{a_M^2 (T_c-T_F)^2}\left( \frac{g_{M \varepsilon}}{C_{0}}+2\gamma  \right) \sigma  ,\label{ssucpar2}\\
\chi_\perp(T_{c}) &\approx&  \frac{1}{a_M (T_c-T_F)} - \frac{2}{\sqrt{3}} \frac{1}{a_M^2 (T_c-T_F)^2}\left( \frac{g_{M \varepsilon}}{C_{0}} -\gamma  \right) \sigma \label{ssucper2},
\end{eqnarray}
where we introduced the notation
\begin{eqnarray}
\gamma = \frac{1}{\sqrt{3}}\frac{ B g_{M Q}}{B^2-C_{3}a_{Q}(T_{c}-T_{Q})},\label{gamma}
\end{eqnarray}
and $T_{c}$ means the transition temperature in the presence of uniaxial pressure given by eq.(\ref{pbs1}). 

If the condition $g_{M \varepsilon}/C_{0} \approx \gamma$ is satisfied accidentally, the term with uniaxial pressure $\sigma$ disappears in the expression of $\chi_\perp(T_{c})$. We interpret the experimental situation in this way, and obtain the susceptibilities as
\begin{align}
\chi_\parallel(T_{c}) &=   \frac{1}{a_M (T_c-T_F)}+\Gamma \cdot \sigma,
&\chi_\perp(T_{c})& =  \frac{1}{a_M (T_c-T_F)} \label{eq:susces1a},
\end{align}
where
\begin{eqnarray}
\Gamma=
-\frac{1}{a_M^2 (T_c-T_F)^2}\frac{2g_{M Q}B}{B^2-a_{Q}(T_c-T_{Q})C_{3}}=-\frac{2\sqrt{3}}{a_M^2 (T_c-T_F)^2}\frac{g_{M \varepsilon}}{C_{0}}.
\end{eqnarray}
We find that with the choice of $ \Gamma=2.85\cdot 10^{5}$$[\mu_{B}^2$$\cdot$GPa$^{-2}]$ the measured susceptibilities $\chi_\parallel(T_{c})$ and $\chi_\perp(T_{c})$ can be fitted at the transition temperature. The result is shown in the right part of Fig.~\ref{fig:4}.
Different anisotropic behavior is realized depending on the parameters $g_{M \varepsilon}$ and $g_{M Q}$.

From the expression (\ref{eq:gibbs3}) we can calculate the magnetic susceptibility for a general direction of the magnetic field with respect to the uniaxial pressure $\sigma\|(001)$.
We obtain the magnetic susceptibility for the magnetic field direction $(h_{x},h_{y},h_{z})$ at the transition temperature as
\begin{eqnarray}
\chi(T_c) =   \frac{1}{a_M (T_c-T_F)}+\left[ a_{1}\left(h_{x}^2+h_{y}^2+h_{z}^2 \right) +a_{2}\frac{1}{2}\left(h_{x}^2+h_{y}^2-2h_{z}^2 \right) \right]\sigma,
\end{eqnarray}
where
\begin{align}
a_{1}& = - \frac{2}{\sqrt{3}}  \frac{1}{a_{M}^2(T_{c}-T_{F})^2} \frac{ g_{M \varepsilon}}{C_{0}},
&a_2 &= \frac{4}{3}  \frac{1}{a_{M}^2(T_{c}-T_{F})^2}\frac{B g_{M Q}}{\left(B^2-C_{3}a_{Q}(T_{c}-T_{Q})\right)}.
\end{align}
We require that in the case of $H\perp \sigma$ with $h_{x}=1$, $h_{y}=h_{z}=0$, the uniaxial pressure term disappears in the magnetic susceptibility. Using this condition $a_{1}+1/2a_{2}=0$, we obtain
\begin{eqnarray}
\chi(T_c)=  \frac{1}{a_M (T_c-T_F)} +3a_{1}h_{z}^2\sigma = \frac{1}{a_M (T_c-T_F)}+\left(3a_{1}\cos^2\theta\right)  \sigma.
\end{eqnarray}

\subsection{Nonlinear effects of magnetic field}

\begin{figure}
\centering
\includegraphics[totalheight=5cm,angle=0]{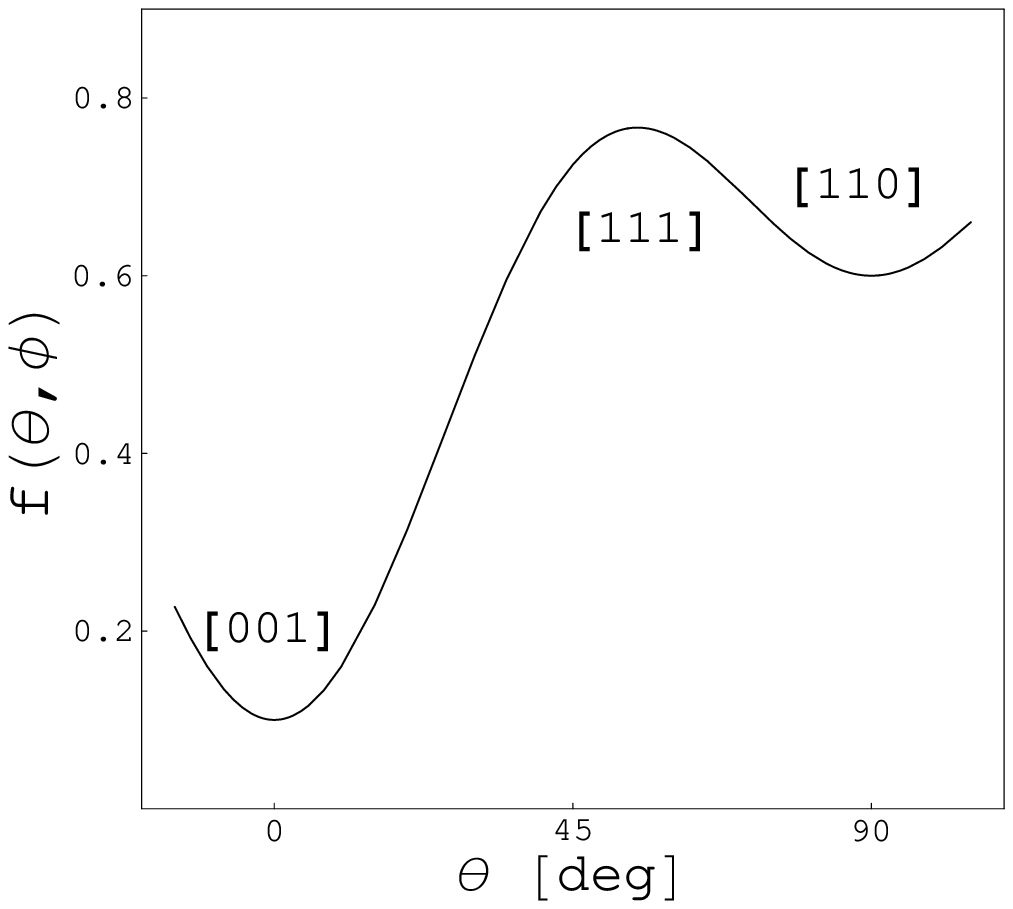}\hspace*{1cm}
\includegraphics[totalheight=5cm,angle=0]{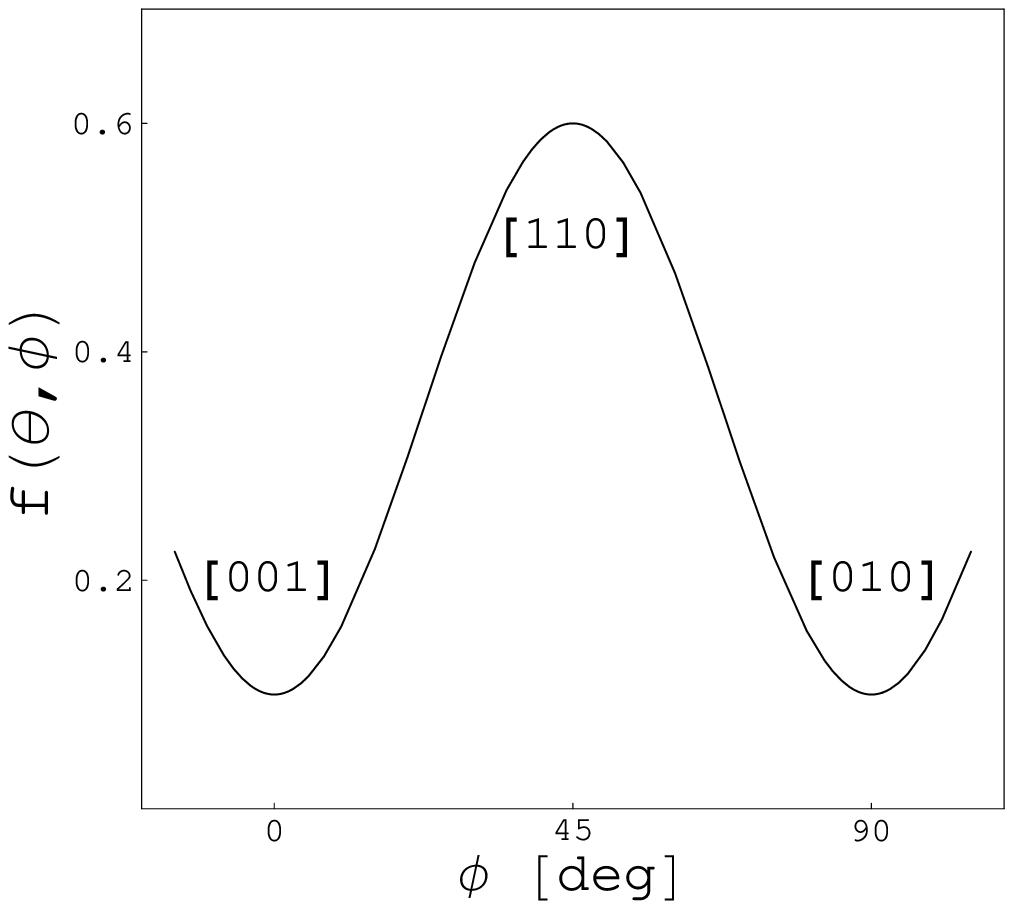}
\caption{Field angle dependence of function $f(\theta,\phi)$ with $\phi=\pi/4$ ({\sl left}) and $\theta=\pi/2$ ({\sl right}), which means the planes [1,-1,0] and [0,0,1], respectively. The parameter values are chosen as $t=1/2$ and $t_{a}=-1$.}\label{fig:1}
\end{figure}

In order to discuss the anisotropy under magnetic field, we have to 
include the fourth order terms of magnetization in ${\cal F}(\psi_{\bf Q}, \vec{M})$.
It is then more convenient to perform the Legendre transformation to magnetic field $\vec{H}$, to obtain 
the free energy expansion as
\begin{eqnarray}
{\cal F}(\psi_{\bf Q},\vec{H})={\cal F}_{0}(\vec{H})+\frac{1}{2}a_{\psi}[T-T_{c}(\vec{H})]\psi_{\bf Q}^2+\frac{1}{4}b_{\psi}\psi_{\bf Q}^4,\label{eq:free1}
\end{eqnarray}
where
\begin{eqnarray}
T_{c}(\vec{H})=T_{0}-
t_{H2}H^2+(t_{4}+t_{4a}h_{4})H^4\label{eq:tc1}
\end{eqnarray}
is the transition temperature in the presence of magnetic field. Here, $t_{H2}$, $t_4$ and $t_{4a}$ are coefficients, and $h_{4}$ is the fourth order cubic invariant $h_4=h_x^4+h_y^4+h_z^4-3/5$, which is common in both cubic $O_{h}$ and tetrahedral $T_{h}$ symmetries. 
The reference part ${\cal F}_{0}(\vec{H})$ has the field dependence similar to $T_{c}(\vec{H})$.

The second order term in the expression of the transition temperature given by eq.(\ref{eq:tc1}) is isotropic, while the
fourth order term carries anisotropy expressed by the function $f(\theta,\phi)\equiv t+t_{a}h_4$, where the field angle $\theta$ and $\phi$ are defined as $(h_x,h_y,h_z)=({\rm sin} \theta{\rm cos}\phi,{\rm sin}\theta{\rm sin}\phi,{\rm cos}\theta)$.
We note that the same function $f(\theta,\phi)$ determines the magnetization anisotropy as well for small magnetic fields.  
Interestingly, the anisotropy given by the function $f(\theta,\phi)$ is independent of the parameter values of the free energy expansion, i.e., the microscopic details of the scalar order.
Figure~\ref{fig:1} shows the field angle dependence of the function $f(\theta,\phi)$ in the $[1,-1,0]$ and $[0,0,1]$ planes. 
We find that for the three principal axes of the magnetic field, namely for the field directions $(001)$, $(110)$ and $(111)$, the following ratio
\begin{eqnarray}
\frac{f[001]-f[111]}{f[110]-f[111]}=4\label{eq:ratio}
\end{eqnarray} 
is held, where we used that $h_{4}[001]=2/5$, $h_{4}[110]=-1/10$ and $h_{4}[111]=-4/15$.
Thus, the ratio (\ref{eq:ratio}) is satisfied for the transition temperature and also for the magnetization as long as the magnetic field is small.
The function $f(\theta,\phi)$ gives an excellent fit for both the observed transition temperature and magnetization in the case of PrFe$_{4}$P$_{12}$ with proper choice for the parameters $t$ and $t_{a}$, which gives strong evidence for scalar order in this compound \cite{sakakibara, utolso}.

\section{Elastic constant}

\subsection{Case of zero magnetic field}

The free energy expansion
in zero magnetic field can be expressed in terms of the strain components as
\begin{eqnarray}
{\cal F}(\varepsilon)=\frac{1}{2}C_{ijkl}\varepsilon_{ij}\varepsilon_{kl},
\end{eqnarray}
where the elastic constant $C_{ijkl}$ can be obtained as $C_{ijkl}=\partial^2 {\cal F}/\partial \varepsilon_{ij}\partial \varepsilon_{kl}$. The connection between the strain and stress components $\sigma_{ij}$ is given by $\sigma_{ij}=\sum C_{ijkl} \varepsilon_{kl}$.

We discuss the properties of the elastic constants $C_{11}-C_{12}$ which is related to the $\Gamma_{3}$ quadrupoles, and $C_{11}+2C_{12}$ which is related to the scalar order parameter. They can be obtained from the free energy expansion (\ref{eq:gibbs3}) as
\begin{align}
C_{11}-C_{12} &=\frac{\partial^2 {\cal F}}{\partial \varepsilon_{u}^2}=\frac{\partial }{\partial \varepsilon_{u}}\left( C_{3}\varepsilon_{u}+BQ_{u}  \right),
&C_{11}+2C_{12} &=\frac{\partial^2 {\cal F}}{\partial \varepsilon_{s}^2}=\frac{\partial }{\partial \varepsilon_{s}}\left( C_{0}\varepsilon_{s}+2g_{\psi \varepsilon}\psi_{\bf Q}^2  \right).
\end{align}
Let us consider first the elastic constant $C_{11}-C_{12}$. The condition $\partial {\cal F}/\partial Q_{u}=0$ gives that $Q_{u}=-(B/\tilde{\alpha}_{Q})\varepsilon_{u}$, where $\tilde{\alpha}_{Q}$ contains $\psi_{\bf Q}^2$ and $\varepsilon_{s}$. 
We set $\psi_{\bf Q}^2$ and $\varepsilon_{s}$
using the equilibrium conditions 
$\partial {\cal F}/\partial \psi_{\bf Q} =0$ and $\partial {\cal F}/\partial \varepsilon_{s} =0$, and obtain
\begin{eqnarray}
C_{11}-C_{12}=C_{3}-B^2\left(\alpha_{Q} -\frac{c_{\psi Q} \alpha_{\psi}}{b_{\psi}} 
+\frac{2g_{\psi \varepsilon}\alpha_{\psi}}{b_{\psi}C_{0} -2g_{\psi \varepsilon}^2}
\left(g_{Q \varepsilon}-\frac{c_{\psi Q}g_{\psi \varepsilon}}{b_{\psi}} \right)\right)^{-1}.\label{fullelc}
\end{eqnarray}
Neglecting the last term of the denominator in eq.(\ref{fullelc}), we obtain
\begin{eqnarray}
C_{11}-C_{12}= C_{3}-\frac{B^2}{\alpha_{Q} -c_{\psi Q} \alpha_{\psi}/b_{\psi}} =C_{11}^{(0)}-C_{12}^{(0)}-\frac{B^2}{\alpha_{Q} -c_{\psi Q} \alpha_{\psi}/b_{\psi}}.\label{elc3}
\end{eqnarray}
For the elastic constant $C_{11}+2C_{12}$ we obtain 
\begin{eqnarray}
C_{11}+2C_{12}=  C_{0}-\frac{2g_{\psi \varepsilon}^2}{b_{\psi}} =  C_{11}^{(0)}+2C_{12}^{(0)}-\frac{2g_{\psi \varepsilon}^2}{b_{\psi}}.\label{elc0}
\end{eqnarray}
In the disordered phase ($\psi_{\bf Q}=0$),  the elastic constant $C_{11} -C_{12}$ is modified as
\begin{eqnarray}
\delta(C_{11} -C_{12})=
-\frac{B^2}{a_Q (T-T_Q)}= 
-\frac{B^2\chi_{Q}}{1-\chi_{Q}\lambda_{Q}},\nonumber
\end{eqnarray}
where we introduced $\delta C_{ij}=C_{ij}-C_{ij}^{(0)}$, and $\chi_{Q}$ is the quadrupolar susceptibility.
At high temperatures it has the property $\delta(C_{11} -C_{12})
\sim 1/T$, which is analogous to the Curie law.

\begin{figure}[t]
\centering
\includegraphics[height=5cm,angle=0]{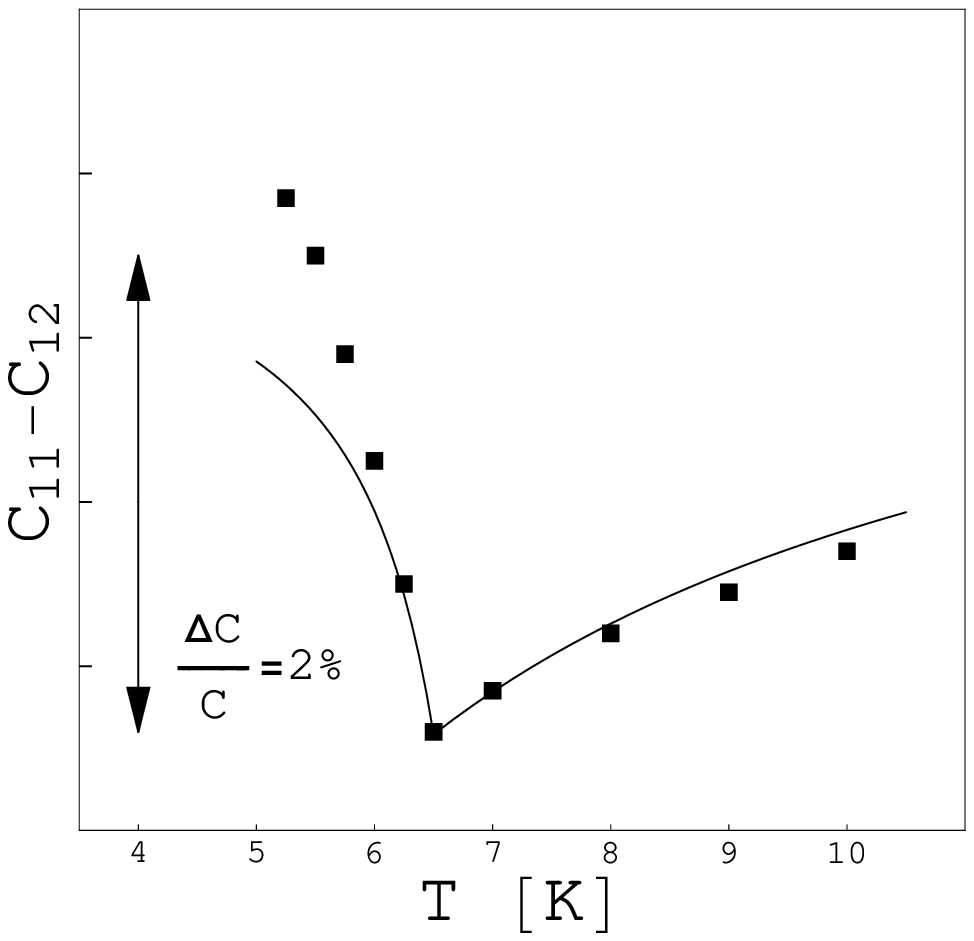}\hspace*{1cm}
\includegraphics[height=5cm,angle=0]{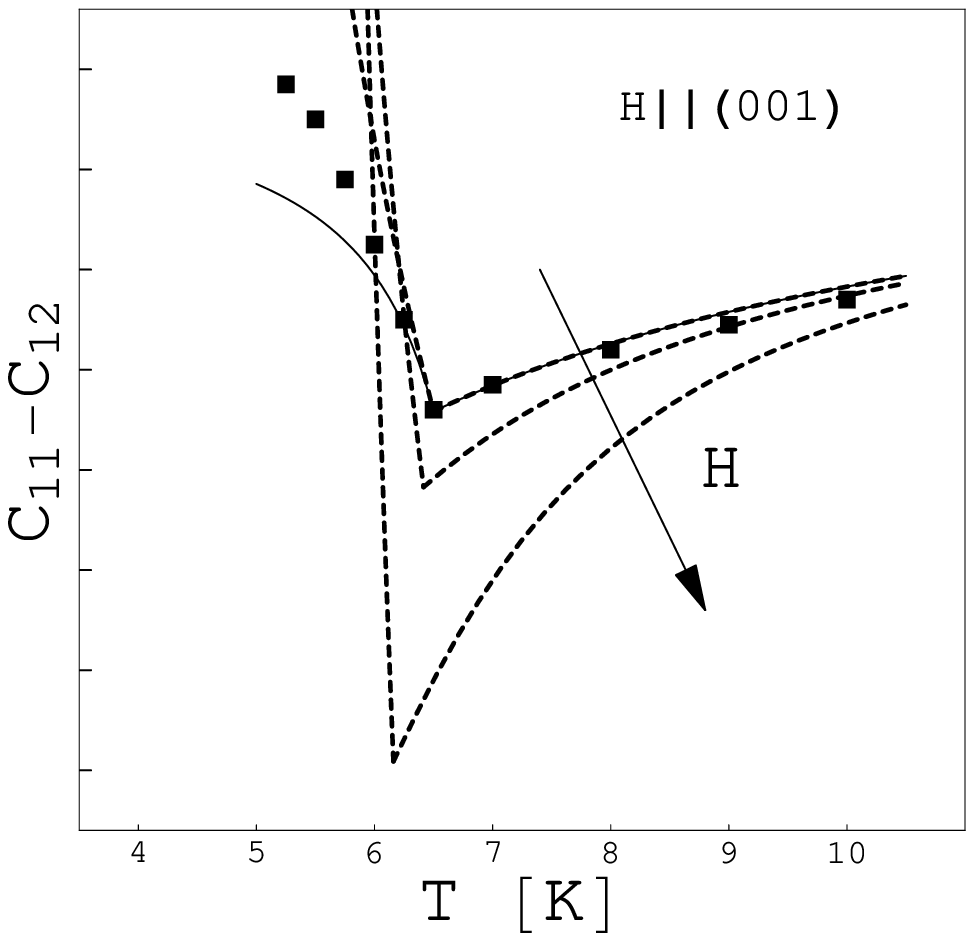}
\hspace*{1cm}
\caption{{\sl Left:} Elastic constant given by the expression (\ref{elc3}) around the transition temperature $T_{0}$ in the case of zero magnetic field.
{\sl Right:} Elastic constant in magnetic fields $H=0, 0.5$ and 1T given by the series expansion (\ref{pceH1}).
Black boxes represent the measured result taken from ref.~\citen{nakanishi2}. The parameter values are given in the text.}\label{fig:3}
\end{figure}

In the ordered phase close to the transition temperature we obtain the 
change 
$\delta C_{ij}$ 
of elastic constant as
\begin{eqnarray}
\delta \left(C_{11}-C_{12} \right) =-\frac{B^2b_{\psi}}{a_Q (T-T_Q)b_{\psi}-c_{\psi Q}a_\psi (T-T_0) }.
\end{eqnarray}
If the condition $\partial  \left(C_{11}-C_{12}\right)/\partial T|_{T_{0}}<0$ is satisfied,
the elastic constant $C_{11}-C_{12}$ has a negative peak at $T=T_0$ due to the scalar order. 
The temperature derivative of $C_{11}-C_{12}$ at $T= T_{0}$ is calculated as
\begin{eqnarray}
\left. \frac{\partial \left(C_{11}-C_{12}\right)}{\partial T}\right|_{T_{0}}
=\left. \frac{\partial \delta \left(C_{11}-C_{12}\right)}{\partial T}\right|_{T_{0}}
=\frac{B^2 \left( a_{Q}b_{\psi} - a_{\psi}c_{\psi Q} \right) }{b_{\psi} a_{Q}^2 (T_{0}-T_{Q})^2},\label{eq:8ce}
\end{eqnarray}
which gives the condition $ a_{\psi}c_{\psi Q} >a_{Q}b_{\psi}$ for the negative peak.

We take the value for $C_3$ from experiment, which gives $C_{11}^{(0)}-C_{12}^{(0)}=70$GPa at $T=4.2$K,\cite{nakanishi1} and choose $T_{Q}=0.4$K. Then we obtain the values for the combinations $B^2/a_{Q}$ and $B^2/c_{\psi Q}$ from the fit of the experimental data\cite{nakanishi2} as $B^2/a_{Q}=10$[GPa$\cdot$K], $B^2/c_{\psi Q}=2.2$[GPa]. The calculated elastic constant $C_{11}-C_{12} $ around the transition temperature $T_{0}$ can be seen in the left part of Fig.~\ref{fig:3}.

Around the phase transition we make series expansion of expressions (\ref{elc3}) and (\ref{elc0}) in terms of $(T-T_0)$, and obtain the 
change 
$\delta C_{ij}$ 
of elastic constants as
\begin{eqnarray}
\delta(C_{11} -C_{12}) &\approx &  
- \frac{B^2}{a_Q (T-T_Q)} \left[  1+
\frac{a_{\psi }
c_{\psi Q}}
{b_{\psi}a_{Q}} \cdot
\frac{T-T_{0} }{T-T_{Q} }
\right]
,\label{eq:invel7}\\
\delta(C_{11} +2C_{12} )&\approx & 
 - \frac{2 g_{\psi \varepsilon}^2}{b_{\psi}}
.\label{eq:invel8}
\end{eqnarray}
In the present mean field theory $\delta \left(C_{11} -C_{12}\right)$ is continuous at $T=T_{0}$, and only its temperature derivative has a discontinuity. On the other hand, the elastic constant $C_{11} +2C_{12}$ is discontinuous at $T=T_{0}$, namely $\Delta \left(C_{11} +2C_{12}\right)=\delta \left(C_{11} +2C_{12}\right)_{T_{0}^{-}}-\delta \left(C_{11} +2C_{12}\right)_{T_{0}^{+}} \ne 0$, and the jump $\Delta \left(C_{11} +2C_{12}\right)$ is always negative in the mean field theory.

Using the values for the parameters obtained, the change of the elastic constant $C_{11}-C_{12}$ around the transition is estimated as
\begin{eqnarray}
\frac{\Delta C }{\Delta T}  \approx \Delta   \left. \frac{\partial \left(C_{11}-C_{12}\right)}{\partial T}\right|_{T_{0}}=-\left(\frac{B^2}{a_{Q}}\right)^2 \frac{c_{\psi Q}}{B^2}
\frac{a_{\psi}}{b_{\psi}(T_{0}-T_{Q})^2}\approx 2.3 {\rm GPa}\cdot{\rm K}^{-1}
,\label{eq:5ce}
\end{eqnarray}
which has the same order of magnitude as the inverse of parameter $t_{\sigma 1} $. The latter describes the temperature-uniaxial pressure phase boundary as $t_{\sigma 1} = \Delta T_{c}/\Delta \sigma$ (see eq.(\ref{pbs1})).
We also estimate the relative change in the elastic constant $C_{11}-C_{12}$ around the transition temperature as
\begin{eqnarray}
\frac{1}{C}\frac{\Delta C}{\Delta T}\approx \frac{1}{C_{11}^{(0)}-C_{12}^{(0)}} \Delta   \left( \frac{\partial \left(C_{11}-C_{12}\right)}{\partial T}\right)_{T=T_{0}}\approx 3.2\cdot 10^{-2}\hspace*{0.1cm}{\rm K}^{-1},
\end{eqnarray}
which means a few percent change of the elastic constant around $T_{0}$ compared to the background value $C_{11}^{(0)}-C_{12}^{(0)} $ in accordance with the experimental situation.

\subsection{Case of non-zero magnetic field}

Now we consider the effect of non-zero external magnetic field with direction $(001)$ on the elastic constant $C_{11}-C_{12}$. We use the relation 
\begin{align}
C_{11}-C_{12} &=\frac{\partial^2 {\cal F}}{\partial \varepsilon_{u}^2}=\frac{\partial }{\partial \varepsilon_{u}}\left( C_{3}\varepsilon_{u}+BQ_{u}  \right).
\end{align}
The condition $\partial {\cal F}/\partial Q_{u}=0$ gives the quadrupolar moment as $Q_{u}=-(B/\tilde{\alpha}_{Q})\varepsilon_u$, where $\tilde{\alpha}_{Q}$ now contains also the magnetization $M_{z}\equiv M$ besides $\psi_{\bf Q}$ and $\varepsilon_{s}$.
We set $\psi_{\bf Q}^2$ and $\varepsilon_{s}$ using the equilibrium conditions 
$\partial {\cal F}/\partial \psi_{\bf Q} =0$, $\partial {\cal F}/\partial \varepsilon_{s} =0$, and $H=\tilde{\alpha}_{M}M$.
We make series expansion in terms of $T-T_{0}$ in the vicinity of the transition temperature $T_{0}$, and also in terms of the magnetic field since it appears in the denominator. We obtain the elastic constant as
\begin{eqnarray}
 C_{11}-C_{12} &\approx& C_{3} -\frac{B^2}{\alpha_{Q}} + \frac{B^2}{\alpha_{Q}^2 \alpha_{M}^2} \left(c_{M Q} +\frac{4}{3}\frac{g_{M Q}^2}{\alpha_{M}} \right)H^2- \frac{B^2}{\alpha_{Q}^2}\frac{a_{\psi}c_{\psi Q}}{b_{\psi}}(T-T_{c})\nonumber\\
&+&  \frac{2B^2}{\alpha_{Q}^2\alpha_{M}^2}\frac{a_{\psi}}{b_{\psi}} \left[\left(\frac{c_{\psi M}}{\alpha_{M}}+
\frac{c_{\psi Q}}{\alpha_{Q}} \right)\left(c_{M Q}-\frac{c_{\psi M}c_{\psi Q}}{b_{\psi}} \right) +
\frac{2 g_{M Q}^2}{\alpha_{M}}\left(\frac{c_{\psi M}}{\alpha_{M}}+\frac{2}{3}
\frac{c_{\psi Q}}{\alpha_{Q}} \right) \right]
(T-T_{c})H^2,\nonumber\\\label{pceH1}
\end{eqnarray}
where $T_{c}$ means the transition temperature in the presence of external magnetic field given by eq.(\ref{pbh2p}).
In the paramagnetic phase the elastic constant reduces to 
\begin{eqnarray}
C_{11}-C_{12} =  C_{3}-\frac{B^2}{a_{Q}(T-T_{Q})} + \frac{B^2}{a_{Q}^2(T-T_{Q})^2a_{M}^2(T-T_{F})^2} \left(c_{M Q} +\frac{4}{3}\frac{g_{M Q}^2}{a_{M}(T-T_{F})} \right)H^2,
\label{pceH1p}
\end{eqnarray}
which shows a softening of the elastic constant in magnetic field provided the coupling constant $c_{M Q}$ has a negative sign.
Furthermore, using a proper choice for $c_{M Q}$, the coefficient of term $(T-T_{c})H^2$ in the ordered phase can be negative.
Then the sharpening of the anomaly occurs at $T_{c}$ in magnetic field since the absolute value of the derivative $\partial  \left(C_{11}-C_{12}\right)/\partial T|_{T_{0}}$ increases.
Right part of Fig.\ref{fig:3} shows the behavior of the elastic constant in magnetic field with the parameter value $B^2/c_{M Q}=-0.5$[GPa$\cdot \mu_{\rm B}^2$].

\section{Ehrenfest relations}

\begin{figure}
\centering
\includegraphics[totalheight=5cm,angle=0]{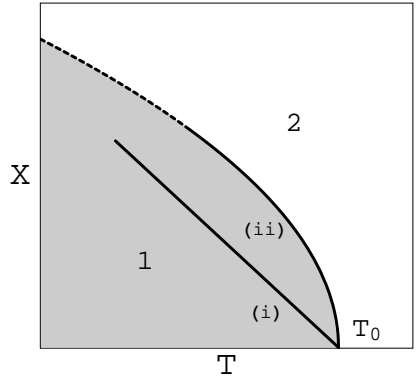}
\caption{Phase boundary in the $T-X$ plane around the transition temperature $T_{0}$.}\label{fig:10}
\end{figure}

Let us regard a continuous phase boundary on the $T-X$ plane, where $T$ is the temperature and $X$ is an arbitrary external quantity such as uniaxial stress, hydrostatic pressure or magnetic field (see Fig.\ref{fig:10}). 
The vicinity to the phase transition $T_{0}$ is considered, where the quantity $X$ is small.
We assume that the phase transition is of second order around $T_{0}$.
The free energy is continuous along the phase boundary, i.e. ${\cal F}(T_{0}+dT,dX)_{1}={\cal F}(T_{0}+dT,dX)_{2}$, where 1 and 2 labels the ordered and paramagnetic phases, respectively.
Furthermore, in the case of a second order transition the first derivatives of the free energy $\partial {\cal F}/\partial T$ and $\partial {\cal F}/\partial X$ are also continuous along the phase boundary.
Two different cases are considered: (i) linear phase boundary given by $T_c=T_{0}-t_{x1}X$ and (ii) quadratic phase boundary given by $T_c=T_{0}-t_{x2}X^2$.
Using the continuity of the free energy and its first derivatives along the phase boundary, we can obtain relations between the anomalies of different thermodynamic quantities at the transition temperature $T_{0}$.
These relations are called Ehrenfest relations.

\subsection{Linear phase boundary}

In the case where the phase boundary has a linear dependence such as $T_c=T_{0}-t_{x1}X$, we obtain the following Ehrenfest relations at the transition temperature $T_{0}$
\begin{eqnarray}
-t_{x1}\frac{\Delta C_s }{T_{0}}-\Delta \frac{\partial^2 {\cal F}}{\partial X \partial T}  &=&  0,\label{eq:erel1a}\\
-t_{x1}^2\frac{\Delta C_s }{T_{0}}-\Delta \frac{\partial^2 {\cal F}}{\partial X^2}  &=&  0,\label{eq:erel1}
\end{eqnarray}
where $C_{s}$ is the specific heat defined as $C_{s}=-T\partial^2 {\cal F}/\partial T^2$.

With $X$ being the uniaxial pressure ($X=\sigma$),
the $T$-$\sigma$ phase boundary was obtained in eq.(\ref{pbs1}) for small values of the uniaxial pressure as
$T_c  = T_{0}-t_{\sigma1} \sigma_{s}$, where $\sigma_{s}=1/\sqrt{3}(\sigma_{xx}+\sigma_{yy}+\sigma_{zz})$
is the hydrostatic pressure.
The quantity $\partial^2 {\cal F}/\partial \sigma_{s} \partial T\equiv \alpha$ in the Ehrenfest relations (\ref{eq:erel1a}) is called thermal expansion coefficient $\alpha$.
Using the equilibrium conditions $\partial {\cal F}/\partial \psi_{\bf Q}=0$ and $\partial {\cal F}/\partial \varepsilon_{s}=\sigma_s$ where the free energy ${\cal F}$ is given by eq.(\ref{eq:gibbs3}), 
the discontinuity in $\alpha$ at the transition is calculated as
\begin{eqnarray}
\Delta \alpha = -\frac{a_{\psi}g_{\psi \varepsilon}}{ b_{\psi}C_{0}-2g_{\psi \varepsilon}^2}.\label{deltk}
\end{eqnarray}
To obtain the anomaly of the specific heat, we calculate the temperature dependence of the free energy (\ref{eq:gibbs3}) in the absence of external fields.
Using the equilibrium conditions $\partial {\cal F}/\partial \psi_{\bf Q}=0$ and $\partial {\cal F}/\partial \varepsilon_{s}=0$, 
we obtain 
\begin{eqnarray}
{\cal F}(T)=-\frac{1}{4}\frac{C_{0}a_{\psi}^2(T-T_{0})^2}{b_{\psi}C_{0}-2g_{\psi \varepsilon}^2},
\end{eqnarray}
which gives the jump of the specific heat at $T=T_{0}$ as
\begin{eqnarray}
\frac{\Delta C_{s}}{T_{0}} = \frac{1}{2}\frac{a_{\psi}^2}{b_{\psi}} +\frac{a_{\psi}^2g_{\psi \varepsilon}}{b_{\psi}\left(b_{\psi}C_{0}-2g_{\psi \varepsilon}^2\right)}=\frac{1}{2}\frac{a_{\psi}^2 C_{0}}{b_{\psi}C_{0}-2g_{\psi \varepsilon}^2}.\label{eq:susspec2}
\end{eqnarray}
Using eqs.(\ref{deltk}) and (\ref{eq:susspec2}) together with 
the expression of $t_{\sigma1} $ given in eq.(\ref{pbs1}), we find 
\begin{eqnarray}
-t_{\sigma 1}\frac{\Delta C_{s}}{T_{0}} - \Delta \alpha=0,
\label{eq:erel1mm}
\end{eqnarray}
which is the special case of (\ref{eq:erel1a}) with $X=\sigma_{s}$.

Since the specific heat shows a jump at the transition in PrFe$_{4}$P$_{12}$\cite{aoki}, we expect discontinuity also of the thermal expansion coefficient.
However, there is no observed discontinuity of the thermal expansion coefficient in ref.\citen{kawana}.
Let us compare with the case of URu$_2$Si$_2$ where experiment with hydrostatic pressure has been performed\cite{elfresh,deVisser}.
In the case of URu$_2$Si$_2$, both the specific heat and thermal expansion coefficient show a jump at the hidden order phase transition, and the Ehrenfest relation (\ref{eq:erel1a}) is satisfied within the experimental error\cite{deVisser}.  
The observed discontinuity of the thermal expansion
is about an order of magnitude 
smaller than the error bar in the experiment for PrFe$_{4}$P$_{12}$ \cite{kawana}.
Both URu$_2$Si$_2$ and PrFe$_{4}$P$_{12}$ have almost the same value in the combination $t_{x1}\Delta C_{s}/T_{0}$ in eq.(\ref{eq:erel1a}).
Therefore, in PrFe$_{4}$P$_{12}$, we expect 
similar magnitude of discontinuity of the thermal expansion coefficient at the transition temperature, which may be observed with improved resolution.

Let us discuss the Ehrenfest relation (\ref{eq:erel1}).
The compliance $S_{11}+2S_{12}$ given by
$-\partial^2 {\cal F}/\partial X^2$ with $X=\sigma_{s}$
in eq.(\ref{eq:erel1}) is related to the elastic constant $C_{11}+2C_{12}$ as $S_{11}+2S_{12}=(C_{11}+2C_{12})^{-1}$.
From expression (\ref{elc0}) we obtain the discontinuity of the elastic constant $C_{11}+2C_{12}$ as
\begin{eqnarray}
 \Delta \left(C_{11}+2C_{12} \right)^{-1}
 =\frac{2g_{\psi \varepsilon}^2}{ C_{0} \left(  b_{\psi}C_{0}-2g_{\psi \varepsilon}^2\right)},
\label{eq:pe4}
\end{eqnarray}
where $\Delta \left(C_{11} +2C_{12}\right)^{-1}= \left(C_{11} +2C_{12}\right)^{-1}_{T_{0}^{-}}- \left(C_{11} +2C_{12}\right)^{-1}_{T_{0}^{+}}$.

Using eqs.(\ref{eq:susspec2}) and (\ref{eq:pe4}) together with 
the expression of $t_{\sigma1} $ given in eq.(\ref{pbs1}), we find 
\begin{eqnarray}
t_{\sigma 1}^2\frac{\Delta C_{s}}{T_{0}} - \Delta \left(C_{11}+2C_{12} \right)^{-1}=0,
\label{eq:erel1m}
\end{eqnarray}
which is the special case of (\ref{eq:erel1}) with $X=\sigma_{s}$,
and relates the jump of the elastic constant to that of the specific heat. 
Since the jump of the latter is finite at the transition temperature in the mean field theory, the inverse elastic constant $(C_{11}+2C_{12})^{-1}$ should also have a finite jump.
However, this jump has so far not been found in the experiment in PrFe$_{4}$P$_{12}$ \cite{nakanishi1}.
We believe that such a jump is indeed present, but is too small to be observed.

\subsection{Quadratic phase boundary}

When the phase boundary has the form $T_c=T_{0}-t_{x2} X^2$, we obtain the following Ehrenfest relation at $T=T_{0}$ in the lowest order
\begin{eqnarray}
 t_{x2}\frac{\Delta C_{s} }{T_{0}} + \frac{1}{2}\Delta \frac{\partial^3 {\cal F}}{\partial X^2\partial T} = 0. \label{eq:erel2}
\end{eqnarray}
Let us consider the case where $X$ is the external magnetic field ($X=H$). 
The temperature-magnetic field phase boundary is obtained as quadratic in the expression (\ref{pbh2}), namely $T_{c}=T_{0}-t_{H2}H^2$. 
From the expression (\ref{fullsus}) we obtain the discontinuity of the temperature derivative of the magnetic susceptibility at $T_{0}$ as
\begin{eqnarray}
\Delta \frac{\partial \chi}{\partial T} &=& \frac{a_{\psi}}{a_M (T_{0}-T_F)^2}\left(\frac{C_{0}c_{\psi M} - 2g_{M \varepsilon} g_{\psi \varepsilon}}{b_{\psi}C_{0}-2g_{\psi \varepsilon}^2}\right).\label{eq:susspec1}
\end{eqnarray}
Using the expression of parameter $t_{H2}$ given by eq.(\ref{pbh2}) together with the jump of the specific heat given in eq.(\ref{eq:susspec2}), we find the relation
\begin{eqnarray}
 t_{\rm H2}\frac{\Delta C_{s} }{T_{0}} - \frac{1}{2}\Delta \frac{\partial \chi}{\partial T}=0, \label{eq:erel5}
\end{eqnarray}
which is the Ehrenfest relation (\ref{eq:erel2}) with $X=H$.

Similarly to the magnetic susceptibility, the elastic constant $C_{11}-C_{12}$ also possesses anomaly at the transition temperature. 
Using the expression (\ref{fullelc}), we find that the discontinuity of the temperature derivative of $C_{11}-C_{12}$ is related to the specific heat jump through the phase boundary parameter $t_{\sigma 2}$ given in eq.(\ref{pbs2}) as
\begin{eqnarray}
  t_{\sigma 2}\frac{\Delta C_{s}}{T_{0}}-\frac{1}{2}\Delta\frac{\partial \left( C_{11}-C_{12}\right)^{-1}
}{\partial T}=0,\label{eq:erel14}
\end{eqnarray} 
which is the Ehrenfest relation (\ref{eq:erel2}) with $X=\sigma_{u}$. 
Combining the expressions (\ref{eq:erel5}) and (\ref{eq:erel14}), we derive the Ehrenfest relation
\begin{eqnarray}
\frac{1}{ t_{\sigma 2}} \Delta\frac{\partial \left( C_{11}-C_{12}\right)^{-1}
}{\partial T} = 
 \frac{1}{t_{H2}}\Delta \frac{\partial \chi}{\partial T},\label{eq:erel15}
\end{eqnarray} 
which relate the anomaly of the elastic constant $C_{11}-C_{12}$ to that of the magnetic susceptibility
at the transition.
We note that parameter $t_{\sigma 2}$ gives the second-order term of the phase boundary relating only to the traceless part of the stress tensor (see eq.(\ref{pbs2})).  
Therefore, it is difficult to extract $t_{\sigma 2}$
from the experimentally observed phase boundary.

Finally, we summarize in Table~\ref{tab:1} the roles of the parameters appearing in the free energy expansion (\ref{eq:gibbs3}) together with the restrictions for the parameters that we found in the Landau analysis.

\begin{table}
\caption{Summary of the parameters appearing in the free energy (\ref{eq:gibbs3}). The restrictions for these parameters obtained from our analysis are also included. The parameter $\gamma$ is defined by eq.({\ref{gamma}}), and $\theta^{*}$ is the Weiss temperature $\theta^{*}=3.5K$\cite{aoki}.
} \label{tab:1}
\centering
\begin{tabular}{|c|c|c|}
\hline {\sl Parameters} & {\sl Relevant quantities} & {\sl Restrictions}  \\\hline\hline
$a_{M}$, $T_{F}$ & $\chi_{+}(T)$ & $a_{M}>0$, $T_{F}=\theta^{*}$ \\\hline
$a_{\psi}$, $c_{\psi M}$ & $\chi_{-}(T)$  & $a_{\psi}c_{\psi M}>a_{M}b_{\psi}$\\
 &  $T$-$H$ phase boundary &   $c_{\psi M}>0$ \\\hline
$C_{11}^{(0)}+2C_{12}^{(0)}$, $a_{\psi}$, $g_{\psi \varepsilon}$  &  $T$-$\sigma$ phase boundary &  $g_{\psi \varepsilon}>0$ \\\hline
 $C_{11}^{(0)}-C_{12}^{(0)}$, $B$, $a_{Q}$, $T_{Q}$ & $\left(C_{11}-C_{12}\right)(H=0,T>T_{0}) $ &  $a_{Q}>0$\\\hline
 $a_{\psi}$, $c_{\psi Q}$ & $\left(C_{11}-C_{12}\right)(H=0,T<T_{0}) $ & $a_{\psi}c_{\psi Q}>a_{Q}b_{\psi}$ \\\hline
 $g_{M Q}$, $g_{M \varepsilon}$ & $\chi_{\parallel}(\sigma)$, $\chi_{\perp}(\sigma)$ &   \\
 & $\chi_{\perp}\approx \chi_{-}(T=T_{c})$ & $g_{M \varepsilon}/C_{0}\approx \gamma$ \\\hline
$c_{M Q}$ & $\left(C_{11}-C_{12}\right)(H) $ & $c_{M Q}<0$ \\\hline 
\end{tabular}
\end{table}

\section{Summary}

In this paper
we have shown that the main properties of PrFe$_{4}$P$_{12}$ can be explained consistently within the scalar order scenario.
We have discussed the effect of a scalar order on magnetic and elastic properties,
and applied the results to the case of PrFe$_{4}$P$_{12}$.
It is found that the anomaly of the magnetic susceptibility and the elastic constant at the transition temperature can be reproduced consistently together with the phase boundaries against to
temperature-magnetic field and temperature-uniaxial pressure. 
Furthermore, we obtain the 
enhancement of the magnetic susceptibility for the uniaxial pressure applied parallel to the direction of the magnetic field, and the almost constant behavior of the susceptibility when the uniaxial pressure is applied perpendicular to the magnetic field.
In our view this behavior is caused by the competition of the isotropic and anisotropic stress dependent terms.
The magnetic susceptibility is calculated also for arbitrary direction of the magnetic field with the uniaxial pressure fixed along $(001)$.
We have studied Ehrenfest-type relations which connect the anomalies of different kinds of thermodynamic quantities at the transition temperature such as specific heat, magnetic susceptibility or elastic constant.
The phenomenological analysis leaves several questions open about PrFe$_{4}$P$_{12}$ such as the microscopic nature of the scalar order parameter, applicability of localized picture for the 4$f$ electrons or the ground state at low temperatures.
Further studies are necessary to clarify the microscopic origin of the scalar order parameter in PrFe$_{4}$P$_{12}$.

\section*{Acknowledgments}

We are grateful to K. Iwasa and T. Sakakibara for sharing their experimental results prior to publication, and 
to Y. Nakanishi and M. Yoshizawa
for useful discussions.


\begin{thebibliography}{999}



\bibitem{aoki}
Y. Aoki, T. Namiki, T. D. Matsuda, K. Abe, H. Sugawara, H. Sato, Phys. Rev. B {\bf 65} (2002) 064446.

\bibitem{tayama}
T. Tayama, J. Custers, H. Sato, T. Sakakibara, H. Sugawara, H. Sato, J. Phys. Soc. Japan {\bf 73} (2004) 3258.

\bibitem{hidaka}
H. Hidaka, I. Ando, H. Kotegawa, T. C. Kobayashi, H. Harima, M. Kobayashi, H. Sugawara, H. Sato, Phys. Rev. B {\bf 71} (2005) 073102. 

\bibitem{namiki}
T. Namiki, Y. Aoki, Y. Yamada, T. D. Matsuda, H. Sugawara, H. Sato, Physica B {\bf 312-313} (2002) 825.

\bibitem{hao2}
L. Hao, K. Iwasa, M. Nakajima, D. Kawana, K. Kuwahara, M. Koghi,
H. Sugawara, T. D. Matsuda, Y. Aoki and H. Sato, Acta Physica
Polonica B {\bf 34} (2003) 1113.

\bibitem{iwasa3}
K. Iwasa, Y. Watanabe, K. Kuwahara, M. Kohgi, H. Sugawara, T. D. Matsuda, Y. Aoki, H. Sato, Physica B {\bf 312-313} (2002) 834.



\bibitem{kikuchi2}
J. Kikuchi, M. Takigawa, H. Sugawara, H. Sato, Physica B {\bf
359-361} (2005) 877.

\bibitem{nakanishi1}
Y. Nakanishi, T. Simizu, M. Yoshizawa, T. Matsuda, H. Sugawara, H. Sato, Phys. Rev. B {\bf 63} (2001) 184429.


\bibitem{hao1}
L. Hao: Thesis.

\bibitem{kikuchi1}
J. Kikuchi, M. Takigawa, H. Sugawara, H. Sato, J. Phys. Soc. Japan {\bf 76} (2007) 043705.


\bibitem{sakai}
O. Sakai, J. Kikuchi, R. Shiina, H. Sato, H. Sugawara, M. Takigawa, H. Shiba, J. Phys. Soc. Japan {\bf 76} (2007) 024710.

\bibitem{sakakibara}
H. Sato, T. Sakakibara, T. Tayama, T. Onimaru, H. Sugawara, H. Sato, 
J. Phys. Soc. Japan {\bf 76} (2007) 064701.



\bibitem{utolso}
A. Kiss and Y. Kuramoto, J. Phys. Soc. Japan {\bf 75} (2006) 103704.

\bibitem{saha}
S. R. Saha: Thesis.


\bibitem{matsuda}
T. D. Matsuda, S. R. Saha, T. Namiki, H. Sugawara, Y. Aoki, H. Sato, J. Phys. Soc. Japan {\bf 71} (2002) Suppl. 246.



\bibitem{nakanishi2}
Y. Nakanishi, M. Yoshizawa, T. Yamaguchi, H. Hazama, Y. Nemoto, T. Goto, T. D. Matsuda, H. Sugawara, H. Sato, J. Phys.: Condens Matter {\bf 14} (2002) L715.


\bibitem{kawana}
D. Kawana, K. Kuwahara, M. Sato, M. Takagi, Y. Aoki, M. Kohgi, H. Sato, H. Sagayama, T. Osakabe, K. Iwasa, H. Sugawara, 
J. Phys. Soc. Japan {\bf 75} (2006) 113602.

\bibitem{elfresh}
M. W. McElfresh, J. D. Thompson, J. O. Willis, M. B. Maple, T. Kohara, M. S. Torikachvili, Phys. Rev. B {\bf 35} (1987) 43.


\bibitem{deVisser}
A. de Visser, F. E. Kayzel, A. A. Menovsky, J. J. M. Franse, J. van den Berg, G. J. Nieuwenhuys, Phys. Rev. B, {\bf 34} (1986) 8168.


\end{thebibliography}
\end{document}